\documentclass[a4paper,11pt]{article}

\pdfoutput=1 

\usepackage{jcappub} 
\usepackage[T1]{fontenc} 

\usepackage{multirow} 
\usepackage{amsmath}
\usepackage{amssymb}
\usepackage{latexsym}
\usepackage{subcaption}

\title{Robustness of predicted CMB fluctuations in Cartan $F(R)$ gravity}

\author[1,2,3]{Tomohiro Inagaki }
\author[4]{Hiroki Sakamoto}
\author[3]{Masahiko Taniguchi}

\affiliation[1]{Information Media Center, Hiroshima University, Higashi-Hiroshima, 739-8521,Japan}
\affiliation[2]{Core of Research for the Energetic Universe, Hiroshima University, Higashi-Hiroshima, 739-8526, Japan}
\affiliation[3]{Graduate School of Advanced Science and Engineering, Hiroshima University, Higashi-Hiroshima, 739-8526, Japan }
\affiliation[4]{Research Institute for Nanodevices, Hiroshima University, 1-4-2, Kagamiyama, Higashi-Hiroshima, 739-8527, Japan}

\emailAdd{inagaki@hiroshima-u.ac.jp}
\emailAdd{h-sakamoto@hiroshima-u.ac.jp}
\emailAdd{masa-taniguchi@hiroshima-u.ac.jp}

\abstract{
We investigated the cosmology of $F(R)$ gravity rebuilt with the Cartan formalism.
This is called Cartan $F(R)$ gravity.
The well-known $F(R)$ gravity has been introduced to extend the standard cosmology, e.g., to explain the cosmological accelerated expansion as inflation.
Cartan $F(R)$ gravity is based on the Riemann-Cartan geometry.
The curvature $R$ is separated into two parts, one is derived from the Levi-Civita connection and the other from the torsion.
Assuming a matter-independent spin connection, we have successfully rewritten the action of Cartan $F(R)$ gravity into the Einstein-Hilbert action and a scalar field with canonical kinetic and potential terms without any conformal transformations.
This feature simplifies the building and analysis of a new model of inflation.
In this paper, we study two models, the power-law model, and the logarithmic model, and evaluate fluctuations in the cosmological microwave background (CMB) radiation.
We found robust CMB fluctuations via analytical computation 
and confirmed this feature through numerical calculations.}

\begin{document}
\maketitle

\section{Introduction}
Inflation is a paradigm to investigate high-energy physics
beyond $\Lambda$CDM and the standard model.
A number of models explaining inflation have been proposed,
which involve extending the gravity sector and/or adding new matter with
minimal or non-minimal coupling to gravity~\cite{Bassett:2005xm,Maurya:2023soo,Santos:2023nhs,Taghavi:2023ptn}.
One of the most famous extended gravity models is the Starobinsky model~\cite{Starobinsky:1979ty,Starobinsky:1980te}.
By applying a suitable conformal transformation,
that model becomes described by a scalar-tensor theory.
The scalar field has the potential to involve a flat plateau for a large value.
Such potential energy contributes to the accelerating expansion of the universe,
while it dominates the energy density of the universe.
As is well known, the accelerating expansion in the early universe
can solve the horizon and flatness problems~\cite{Sato:1980yn,Guth:1980zm}.

In general, the modification of the Einstein-Hilbert action to
an arbitrary function of the Ricci scalar can be rewritten as an equivalent
scalar-tensor theory through a conformal transformation~\cite{Nojiri:2006ri,Nojiri:2010wj,Nojiri:2017ncd,jordan1955schwerkraft}.
It should be noticed that there is some discussion about the equivalence of physics before and after the conformal transformation~\cite{Catena:2006bd,Steinwachs:2011zs,Kamenshchik:2014waa,Hamada:2016onh}.

Cartan $F(R)$ gravity is an extended model of general relativity (GR), in which the Einstein-Hilbert term is replaced by a function of the curvature scalar to be defined in Sec.~\ref{section::two}. 
We denote the curvature scalar as $R$ instead of the Ricci scalar for simplicity~\cite{Inagaki:2022blm}.
The model $F(R) = R$ has been called Einstein-Cartan-Kibble-Sciama (ECKS) theory
since the 1960s~\cite{SCIAMA:1964rmp,Kibble1961jmp}.
ECKS theory is still actively studied and applied to cosmological problems~\cite{Boehmer:2008ah,Poplawski:2010kb,Magueijo:2012ug,Shaposhnikov:2020frq,Shaposhnikov:2020gts,Iosifidis:2021iuw,Cabral:2021dfe,Piani:2022gon}.

A feature of Cartan $F(R)$ gravity is that the torsion does not vanish~\cite{Montesinos:2020pxv}.
The curvature scalar $R$ is then divided into the usual part obtained from the Levi-Civita connection and an additional part obtained from the torsion.
The non-vanishing torsion is a common feature of the Palatini approach to modified gravity and more general metric-affine geometry.
Several works have been done on the basic properties of metric-affine $F(R)$ gravity, including applications to cosmology~\cite{Sotiriou:2006mu,Sotiriou:2006qn,Iosifidis:2018zjj}.
It was found that metric-affine $F(R)$ and Palatini $F(R)$ gravity are rewritten forms from a certain class of Brans-Dicke type scalar-tensor theories after conformal transformations~\cite{Capozziello:2007tj,Capozziello:2008yx,Sotiriou:2009xt,Capozziello:2009mq,Olmo:2011uz}.
However, the conformal transformation is not necessary to rewrite the Cartan $F(R)$ gravity into an equivalent scalar-tensor theory~\cite{Inagaki:2022blm}.
The aims of this study are to propose a model of the Cartan $F(R)$ gravity consistent with Planck 2018 results and predict the CMB fluctuations.

We organize this paper as follows,
in Sec.~\ref{section::two} we briefly introduce the Cartan formalism and
Cartan $F(R)$ gravity.
We employ the standard slow-roll scenario and calculate the CMB fluctuations in the Cartan $F(R)$ gravity.
In the slow-roll scenario, the evolution of spacetime is characterized by the slow-roll parameters. 
We formulate these parameters and the e-folding number in the Cartan $F(R)$ gravity in Sec.~\ref{section:inflation}.
In Sec.~\ref{section::model_analysis}, we consider the power-law and logarithmic models and calculate the power spectrum, the spectral index, the tensor-scalar ratio, and the running spectral index.
It is found that these model predictions are robust under the variation of the model parameters.
The robustness of the CMB fluctuations is confirmed by numerical calculations in Sec.~\ref{section::numerical-result}.

In Sec.~\ref{section::reheating}, we demonstrate reheating processes in several models of the Cartan $F(R)$ gravity.
Finally, we give some concluding remarks.

\section{Cartan $F(R)$ gravity}
\label{section::two}
We start by reformulating Cartan $F(R)$ gravity on the Riemann-Cartan geometry described by the vierbein ${e^i}_\mu$ and the spin connection ${\omega^{ij}}_\nu$.
The vierbein connects the curved metric $g_{\mu\nu}$ and flat one $\eta_{ij}$ with,
\begin{align}\label{Eq:Definition of tetrad}
g_{\mu\nu}=\eta_{ij}{e^i}_\mu{e^j}_\nu.
\end{align}
Since Cartan $F(R)$ is a natural extension of the conventional $F(R)$ gravity,
we expect the additional contribution to GR can be described as a scalar field theory.
This situation is similar to the conventional $F(R)$ gravity, but no conformal transformation is required.
However, this situation can avoid the difficulty of physical quantities being frame dependent~\cite{Inagaki:2022blm}.

The action of Cartan $F(R)$ gravity is defined by replacing the curvature scalar $R$ 
in Einstein-Cartan theory with a general function $F(R)$,
\begin{align}\label{eqs:action FR}
S=\int d^4xe\left( \frac{{M_{\rm Pl}}^2}{2}F(R)+\mathcal{L}_{\rm{m}}\right),
\end{align}
where $M_{\rm Pl}$ indicates the Planck scale and a volume element is given by the determinant of the vierbein, $e$.
The curvature scalar is expressed by the spin connection and the vierbein,
\begin{align*}
R
= {e_i}^\mu{e_j}^\nu {R^{ij}}_{\mu\nu}(\omega,\partial\omega)
= {e_i}^\mu{e_j}^\nu\left[\partial_{\mu}{\omega^{ij}}_\nu-\partial_{\nu}{\omega^{ij}}_\mu+{\omega^i}_{k\mu}{\omega^{kj}}_\nu-{\omega^i}_{k\nu}{\omega^{kj}}_\mu\right].
\end{align*}

In Cartan geometry, a geometric tensor ${T^{\rho}}_{\mu\nu}$ called torsion arises,
\begin{align*}
{T^{\rho}}_{\mu\nu}\equiv{\Gamma^\rho}_{\mu\nu}-{\Gamma^\rho}_{\nu\mu}.
\end{align*}
Where we have expressed the Affine connection as ${\Gamma^\rho}_{\mu\nu}= {e_a}^\rho D_\nu {e^a}_\mu$ and $D_\nu$ is the covariant derivative for the local Lorentz transformation
\begin{align*}
D_\nu {e^k}_\mu=\partial_\nu {e^a}_\mu+{{\omega^k}_{l\nu}}{e^l}_\mu.
\end{align*}
The Affine connection is not necessary invariant under the replacement of the lower indices, ${\Gamma^\rho}_{\mu\nu}\neq{\Gamma^\rho}_{\nu\mu}$.
Assuming that the matter field is spin connection independent, the torsion is represented by the derivative of $F(R)$ and the vierbein from Ref\cite{Inagaki:2022blm};
\begin{align}\label{eqs:torsion of FR}
{T^k}_{ij}=\frac{1}{2}({\delta^k}_j{e_i}^\lambda-{\delta^k}_i{e_j}^\lambda)\partial_\lambda\ln F'(R).
\end{align}
It should be noted that the torsion vanishes in Einstein-Cartan theory, $F(R)=R$.
Non-vanishing torsion can be obtained by extending $F(R)$.
We extract it from the curvature scalar,
\begin{align}\label{eqs:Ricci scalar into non and torsion2}
R=R_E+T-2{\nabla_E}_\mu T^\mu,
\end{align}
where the subscript $E$ in $R_E$ and $\nabla_E$ stands for the Ricci scalar and the covariant derivative given by the Levi-Civita connection.
$T_\mu$ represents the torsion vector $T_\mu={T^\lambda}_{\mu\lambda}$ and the torsion scalar $T$ is defined to contract the torsion and torsion vector as
\begin{align*}
T=\frac{1}{4}T^{\rho\mu\nu}T_{\rho\mu\nu}-\frac{1}{4}T^{\rho\mu\nu}T_{\mu\nu\rho}-\frac{1}{4}T^{\rho\mu\nu}T_{\nu\rho\mu}-T^\mu T_\mu.
\end{align*}
Thus, the curvature scalar is divided into two parts, $R_E$ and an additional part derived from the torsion.
Substituting Eq.(\ref{eqs:torsion of FR}) into Eq.(\ref{eqs:Ricci scalar into non and torsion2}),
the additional part is represented as
\begin{align}\label{eqs:Ricci scalar to einstein and torsion}
R=R_E-\frac{3}{2}\partial_{\lambda}\ln F'(R)\partial^{\lambda}\ln F'(R)-3 \nabla_E^2 \ln F'(R).
\end{align}

Below we consider a class of Cartan $F(R)$ gravity expressed as $F(R)=R+f(R)$.
The canonical scalar $\phi$ is introduced and defined as
\begin{align}\label{eqs:def scalar field}
\phi\equiv-\sqrt{\frac{3}{2}}{M_{\rm Pl}}\ln F'(R).
\end{align}
Substituting Eq.(\ref{eqs:def scalar field}) into Eq.(\ref{eqs:Ricci scalar to einstein and torsion}), 
the gravity part of the action is rewritten
to be the Einstein-Hilbert term and the scalar field.
\begin{align}
S \ni \int d^4x e\frac{{M_{\rm Pl}}^2}{2}\left(R+f(R)\right)
= \int d^4x e\left(\frac{{M_{\rm Pl}}^2}{2}R_E-\frac{1}{2}\partial_{\lambda}\phi \partial^{\lambda}\phi-V(\phi)\right).
\label{eqs:action for R and scalar}
\end{align}
We assume that $\ln F'(R)$ vanishes at a distance,
consequently the last term in \eqref{eqs:Ricci scalar to einstein and torsion} is a total derivative and can be omitted.
The potential, $V(\phi)$, is defined by,
\begin{align} \label{eqs:potential for phi in general}
V(\phi) \equiv -\frac{{M_{\rm Pl}}^2}{2}\left. f(R)\right|_{R=R(\phi)}.
\end{align}
The potential is expressed as a function of the scalar field $\phi$ through $R=R(\phi)$ by solving Eq.(\ref{eqs:def scalar field}).

Thus, we have derived a scalar-tensor theory (\ref{eqs:action for R and scalar}) without any conformal transformations.
It should be noticed that the potential $V(\phi)$ is different from the one in the scalar-tensor theory obtained from conventional $F(R)$ gravity after the conformal transformation~\cite{Nojiri:2017ncd}.
This is acceptable because various potentials can be obtained from Cartan $F(R)$ models.
As an example, the same potential of the Starobinky model can be derived from a model with $f(R)=-R^2$ in Cartan $F(R)$ gravity;
although this model has an opposite sign of the Starobinky model~\cite{Inagaki:2022blm}.

\section{Slow-roll inflation}
\label{section:inflation}
Next, we consider slow-roll inflation in Cartan $F(R)$ gravity.
Slow-roll inflation is a standard scenario of the early-time expansion of the universe.
A single scalar field, called the inflaton, provides energy for spacetime expansion.
In Cartan $F(R)$ gravity, the scalar field, $\phi$ can be identified as the inflaton.  Thus, it can play a crucial role in inflation.

In the slow-roll scenario, inflation is controlled by the slow-roll parameters, $\varepsilon_H \equiv -\dot{H}/H^2$, $\eta_H = \frac{1}{2} \frac{\ddot{H}}{H\dot{H}}$, and $\xi_H$; where $H$ is the Hubble's parameter.
These parameters can be rewritten by the inflaton potential under the slow-roll approximation.
We denote the rewritten form of the parameters as $\varepsilon_V$, $\eta_V$, and $\xi_V$.
Inflation lasts during $\varepsilon_V < 1$ and ends when $\varepsilon_V = 1$.

In our model, the potential is described as a function of $R$, and the slow-roll parameters are given by
\begin{align}
    \label{epsilonv::slow-roll}
    \varepsilon_V
    &= {
        \frac{1}{3} \Big(\frac{F'}{F''}\Big)^2
        \Big(\frac{f'}{f}\Big)^2
    },
    \\
    \label{etav::slow-roll}
    \eta_V
    &= \frac{2}{3} \Big(\frac{F'}{F''}\Big)^2
    \frac{f'}{f} \Big\{
        \frac{f''}{f'} + \frac{F''}{F'}
        - \frac{F'''}{F''}
    \Big\},
    \\
    \label{xiv::slow-roll}
    \xi_V
    &= {
        \frac{4}{9} \Big(\frac{F'}{F''}\Big)^4
        \Big(\frac{f'}{f}\Big)^2 \Big\{
            \frac{f'''}{f}
            + \frac{3}{2} \frac{f'}{f} \frac{F''}{F'} \Big(1 - \frac{F'F'''}{(F'')^2}\Big)
            + \Big(\frac{F''}{F'}\Big)^2
            + 3 \Big(\frac{F'''}{F''}\Big)^2
            - 3 \frac{F'''}{F'}
            - \frac{F''''}{F''}
        \Big\}
    }.
\end{align}
The e-folding number $N$ of inflation is represented as
\begin{align}
    N
    = \frac{3}{2} \int^{R_*}_{R_\text{end}} dR \Big(\frac{F''}{F'}\Big)^2
    \frac{f}{f'},
    \label{eFoldingNumber::generalForm}
\end{align}
where $R_*$ and $R_\text{end}$ are values when the inflation begins and ends.
The latter value is given from the condition, $\varepsilon_V = 1$. 
The former is evaluated to obtain the suitable e-folds $N = 50$--$60$,
which is required to solve the horizon and flatness problems.

Quantum fluctuations of the inflaton can induce the initial value of the curvature perturbation.
This is characterized by the power spectrum $A_s$, the spectral index $n_s$, and the running spectral index $\alpha_s$,
\begin{align}
 \label{As::Analytic}
    A_s
    &\equiv
        -\frac{f}{16 \pi^2 M_\text{Pl}^2} \Big(\frac{F''}{F'}\Big)^2
        \Big(\frac{f}{f'}\Big)^2,
    \\ \label{ns::Analytic} n_s
        &\equiv 1 - 6\varepsilon_V + 2 \eta_V,
    \\ \label{alphas::Analytic}
    \alpha_s
        &\equiv -24\varepsilon_V^2 + 16 \varepsilon_V \eta_V - 2 \xi_V.
\end{align}
Under the slow-roll approximation, primordial gravitational waves are predicted.
The ratio of the power spectrum of primordial gravitational waves and scalar field is called the tensor-to-scalar ratio.
That ratio is evaluated using the slow-roll parameter,
\begin{align} \label{r::Analytic}
    r = 16\varepsilon_V.
\end{align}
Any predictions of these inflationary parameters $n_s$, $\alpha_s$, and $r$ should satisfy the constraints of Planck 2018~\cite{Planck:2018jri}.

\section{Models and predictions of CMB fluctuation}
\label{section::model_analysis}
Slow-roll inflation is a successful scenario that can explain the early-time accelerating expansion of spacetime.
The soundness of inflation models is determined by their consistency with observations of the CMB fluctuations.
It has been empirically found that a model with a flat region in the inflaton potential can be adjusted to satisfy the constraints of the CMB fluctuations and predict a small tensor-to-scalar ratio.

\subsection{Power-law model}
\begin{figure}
    \centering
    \includegraphics[width=0.7\linewidth]{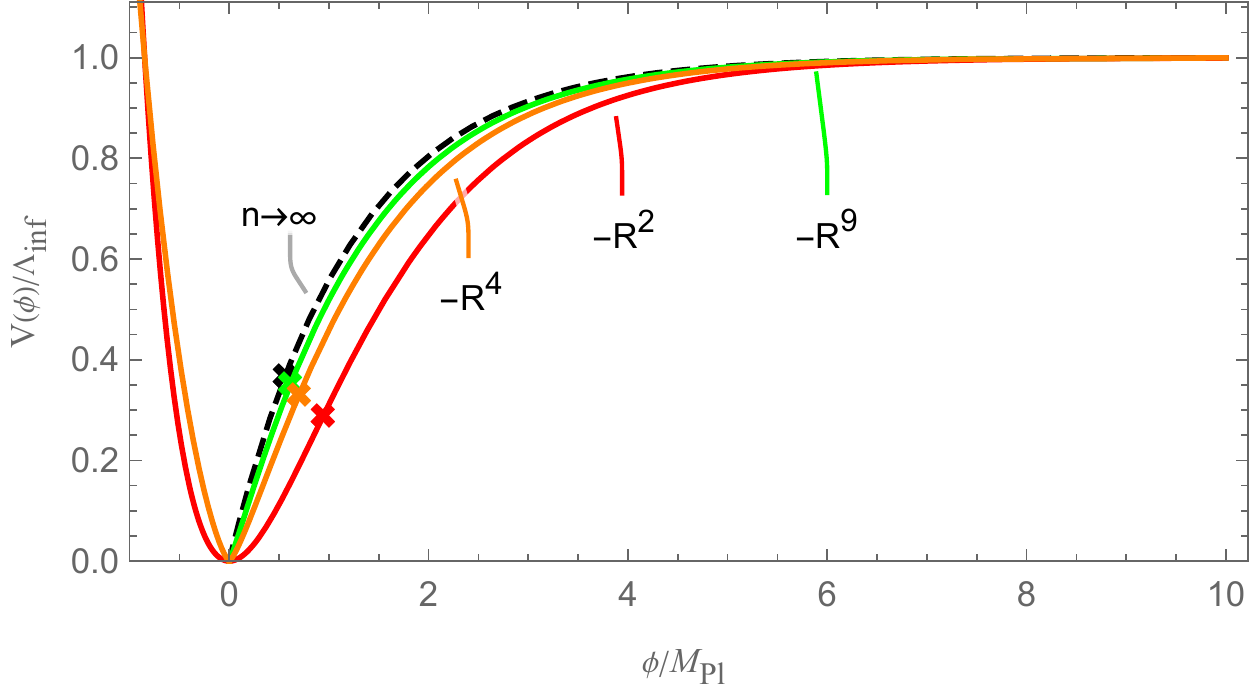}
    \caption{The inflaton potentials of the power-law model \eqref{Model::PowerLaw}.
    The Red, Orange, and Green lines show the potential at $n=2$, $4$, $9$.
    The dashed black line draws the potential of the power-law model with $n\to\infty$.
    Cross marks($\times$) show the end of inflation, the point at which the slow-roll parameter becomes one, $\epsilon_V=1$.}
   \label{Model::Powerlaw::Potential}
\end{figure}

Reference \cite{Inagaki:2022blm} found that the predictions in an $R^2$ model coincides with the predictions of Starobinsky's model.
$R_E^n$ with $n>1$ has been investigated in the conventional $F(R)$ gravity~\cite{Motohashi:2014tra,Inagaki:2019hmm}.
The scalaron potential obtained after the conformal transformation is unstable if $n > 2$, then fine-tuning is required to obtain a suitable e-folding value $N$.
Thus, it is necessary for $n\sim2$ to satisfy observation constraints.
Whereas, in our framework of Cartan $F(R)$ gravity the power-law model,
\begin{align}
    f(R) = -\gamma R^n,\ (n > 1),
    \label{Model::PowerLaw}
\end{align}
introduces a stable potential from Eq.~\eqref{eqs:potential for phi in general}.
Our motivation now is to investigate the higher derivative model of Eq.~\eqref{Model::PowerLaw}.
As is shown in Fig.~\ref{Model::Powerlaw::Potential}, the potential for the power-law model with $n>2$ has a flat plateau.
Consequently, the slow-roll scenarios can be adapted to this model and the quasi-de Sitter expansion is realized around a flat region.

The e-folding number of this model can be analytically obtained as 
\begin{align}
    N = \frac{3}{2} \frac{n-1}{n}
        \Big[\ln(1 - n\gamma R^{n-1}) + \frac{1}{1 - n \gamma R^{n-1}}\Big]
        \Big|_{R_\text{end}}^{R_*},
    \label{eFoldingNumber::PowerLaw}
\end{align}
and we can evaluate $R_*$ at the beginning of inflation by
\begin{align}
    R_* = \Big\{
        \frac{1}{n \gamma} \Big(
            1 + \frac{1}{W_{-1}(x)}
        \Big)
    \Big\}^{1/(n-1)},
\end{align}
where $W_{-1}$ is the Lambert's $W$ function, $x = -\exp(-1-\frac{2nN}{3(n-1)})$, and $R_\text{end}$ is integrated into the normalization of $N$.
The constraint of the power spectrum gives the value of coupling constant $\gamma$.

Next, we calculate the inflationary parameters $A_s$, $n_s$, $\alpha_s$, and $r$.
Substituting Eq.~\eqref{Model::PowerLaw} into Eq.~\eqref{As::Analytic} with Eq.~\eqref{eFoldingNumber::PowerLaw}, we obtain 
\begin{align}
    A_s = 
    \frac{\gamma}{16 \pi^2} \Big\{
        \frac{1}{n\gamma} \Big(
            1 + \frac{1}{W_{-1}(x)}
        \Big)
    \Big\}^{n/(n-1)}
    \Big\{
        \frac{n-1}{n} \Big(
            1 + \frac{1}{W_{-1}(x)}
        \Big)
    \Big\}^2.
    \label{As::PowerLaw::Analytic}
\end{align}
The coupling $\gamma$ is estimated to
satisfied the constraint of power spectrum, $\log 10^{10} A_s = 3.044\pm0.014$~\cite{Planck:2018jri}.
The spectral index is found to be
\begin{align}
    n_s = 1 + \frac{4}{3} \Big(\frac{n}{n-1} \frac{1}{1 + W_{-1}(x)}\Big)^2
    - \frac{2(3n-2)}{3n} \Big(\frac{n}{n-1} \frac{1}{1 + W_{-1}(x)}\Big)^2,
    \label{ns::PowerLaw::Analytic}
\end{align}
and the tensor-to-scalar ratio is
\begin{align}
    r = \frac{16}{3} \Big(\frac{n}{n-1} \frac{1}{1 + W_{-1}(x)}\Big)^2.
    \label{r::PowerLaw::Analytic}
\end{align}
Lastly, the running spectral index takes the form,
\begin{align}
  \begin{aligned}
    \alpha_s = -\frac{8}{9} \Big(\frac{n}{n-1} \frac{1}{1 + W_{-1}(x)}\Big)^2
    + \frac{4}{9} \frac{11n - 12}{n} \Big(\frac{n}{n-1} \frac{1}{1 + W_{-1}(x)}\Big)^3
    \\ - \frac{4}{9} \Big(\frac{9n^2 - 22n + 16}{n^2}\Big) \Big(\frac{n}{n-1} \frac{1}{1 + W_{-1}(x)}\Big)^4.
  \end{aligned}
\end{align}
These formulas are simplified at the $n\to\infty$ limit, and we discuss the results in the next section.

\subsection{Logarithmic model}
In the context of quantum field theory (QFT), integrating out the heavy degrees of freedom (d.o.f) modifies the potential of the light scalar field. 
The logarithmic model, which is obtained as
\begin{align}
    f(R) = -\alpha R \ln \Big( 1 + \frac{R}{R_0} \Big),
    \label{Model::Logarithmic}
\end{align}
mimics the one-loop corrections coming from QFT.
Importantly, the logarithmic model deforms the corrections to keep the Einstein-Hilbert action at the weak curvature limit.
Several variations of the logarithmic corrections have been investigated, as shown in Ref.~\cite{Nojiri:2003ni}.

\begin{figure}
     \centering
     \includegraphics[width=0.7\linewidth]{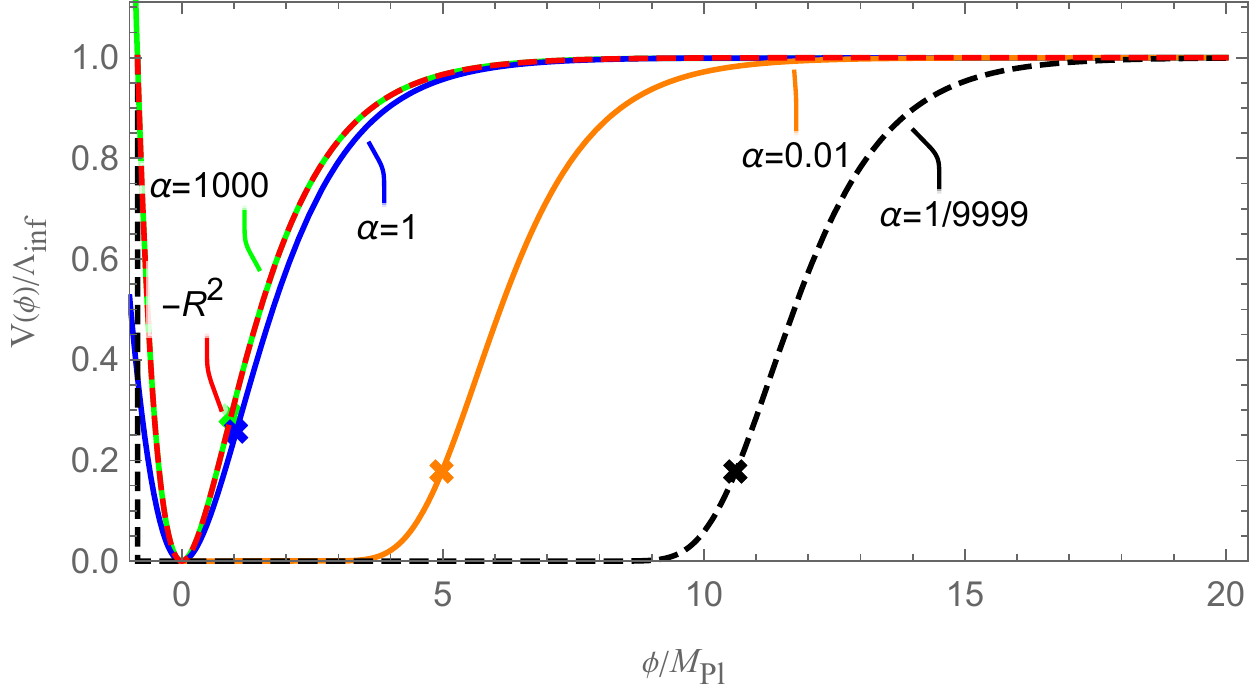}
     \caption{The inflation potential of Eq.~\eqref{Model::Logarithmic} with $\alpha=0.01$, $1$, and $\alpha=1000$. The dashed Red line is Starobinsky potential from the $R^2$ model.
     The dashed black line is the potential with $\alpha=1/9999$.
     Cross marks($\times$) show the end of inflation, the point at which the slow-roll parameter becomes one, $\epsilon_V=1$.}
     \label{Model::Logarithmic::Potential::large}
\end{figure}

\begin{figure}
    \centering
    \includegraphics[width=\linewidth]{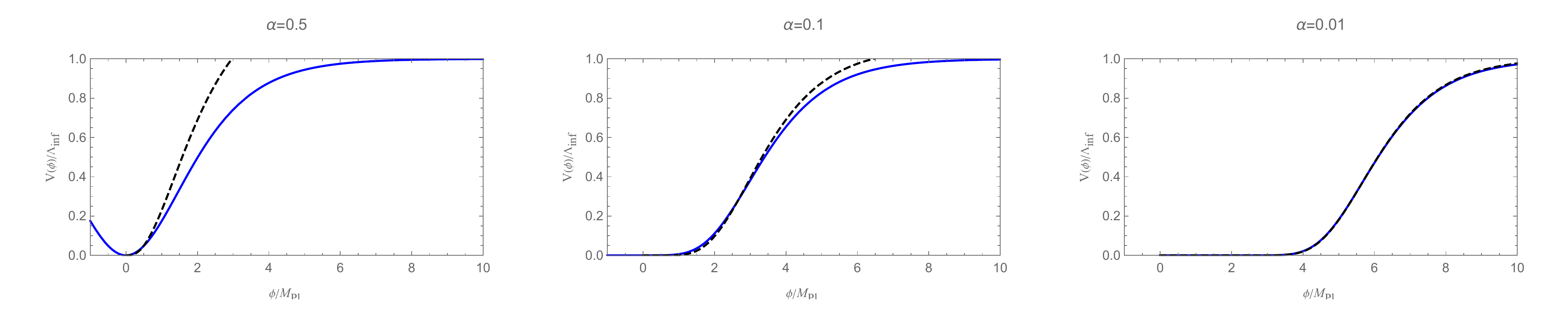}
    \caption{The inflaton potential of Eq.~\eqref{Model::Logarithmic} is the blue line.
    The black line represents the potential of the power-law model with $n=1+\alpha$.}
    
   \label{Model::Logarithmic::Potential::SmallCoupling}
\end{figure}

In Fig.~\ref{Model::Logarithmic::Potential::large},
the inflaton potential of this model has a flat plateau for a large $\phi$ and the quasi-de Sitter expansion is also realized.
Thus, we adopt the large field inflation scenario.
In other words, the inflaton $\phi$ at the beginning of inflation is larger than the Planck scale.
At the limit $\phi\to\infty$ we find from Eq.~\eqref{eqs:def scalar field},
\begin{align*}
    F'(R)=0.
\end{align*}
We can estimate the value of $R$ to solve this equation and the solution is given by
\begin{align}
    R = R_0 (\frac{1}{W(e^{1-1/\alpha})} - 1),
\end{align}
where $W(z)$ is the Lambert's $W$ function.
For a small coupling $\alpha < 1$, $z = \exp(1-1/\alpha)$ is small enough, $z\ll1$.
The Lauran's series of $1/W(z)$ around $z=0$ is
\begin{align}
    \frac{1}{W(z)} = \frac{1}{z} + 1 - \frac{z}{2} + \mathcal{O}(z^2),
\end{align}
and we obtain $R/R_0 \sim e^{1/\alpha} \gg 1$.
Under this assumption, the function $f(R)$ approaches to the power-law model with $n=1+\alpha$,
\begin{align}
    -\alpha R \ln\Big( 1 + \frac{R}{R_0} \Big)
    \sim -\frac{R_0}{e} \Big( \frac{R}{R_0} \Big)^{1+\alpha}.
    \label{Log::approximation::small}
\end{align}
The validity of this approximation can be evaluated to compare the inflaton potential of each model in Fig.~\ref{Model::Logarithmic::Potential::SmallCoupling}.
For a large coupling, the value of $R/R_0$ is less than 1,
because Lambert's $W(z)$ function at $z \sim e$ is almost unity.
In this case, the logarithmic model can be approximated to $f(R) = -(\alpha/R_0) R^2$.
As the coupling $\alpha$ increases, the potential obtained from Eq.~\eqref{Model::Logarithmic} approaches that from Eq.~\eqref{Model::PowerLaw} with $n=2$ (Fig.~\ref{Model::Logarithmic::Potential::large}).
That is, the $\alpha\to\infty$ limit of the logarithmic model is the Starobinsky model, $f(R)=-R^2$.

\section{Numerical result}
\label{section::numerical-result}
We have analytically evaluated CMB fluctuations for Cartan $F(R)$ gravity in the previous sections.
In this section, we numerically calculate the inflationary parameters and show the robustness of the predictions in Cartan $F(R)$ gravity.
We perform the numerical calculations in the following steps:
\begin{enumerate}
\item $R_{\text{end}}$ is found by the condition for the end of the inflation, $\varepsilon_V=1$. 
\item $R_*$ is obtained from Eq.~\eqref{eFoldingNumber::generalForm} with the e-folding number, $N=50,60$.
\item The CMB fluctuations, $n_s$~\eqref{ns::Analytic}, $\alpha_s$~\eqref{alphas::Analytic} and $r$~\eqref{r::Analytic} are estimated from the slow-roll parameters, $\varepsilon_V$~\eqref{epsilonv::slow-roll}, $\eta_V$~\eqref{etav::slow-roll}, $\xi_V$~\eqref{xiv::slow-roll} at $R*$.
\end{enumerate}

\begin{figure}
\centering
      \includegraphics[width=1\linewidth]{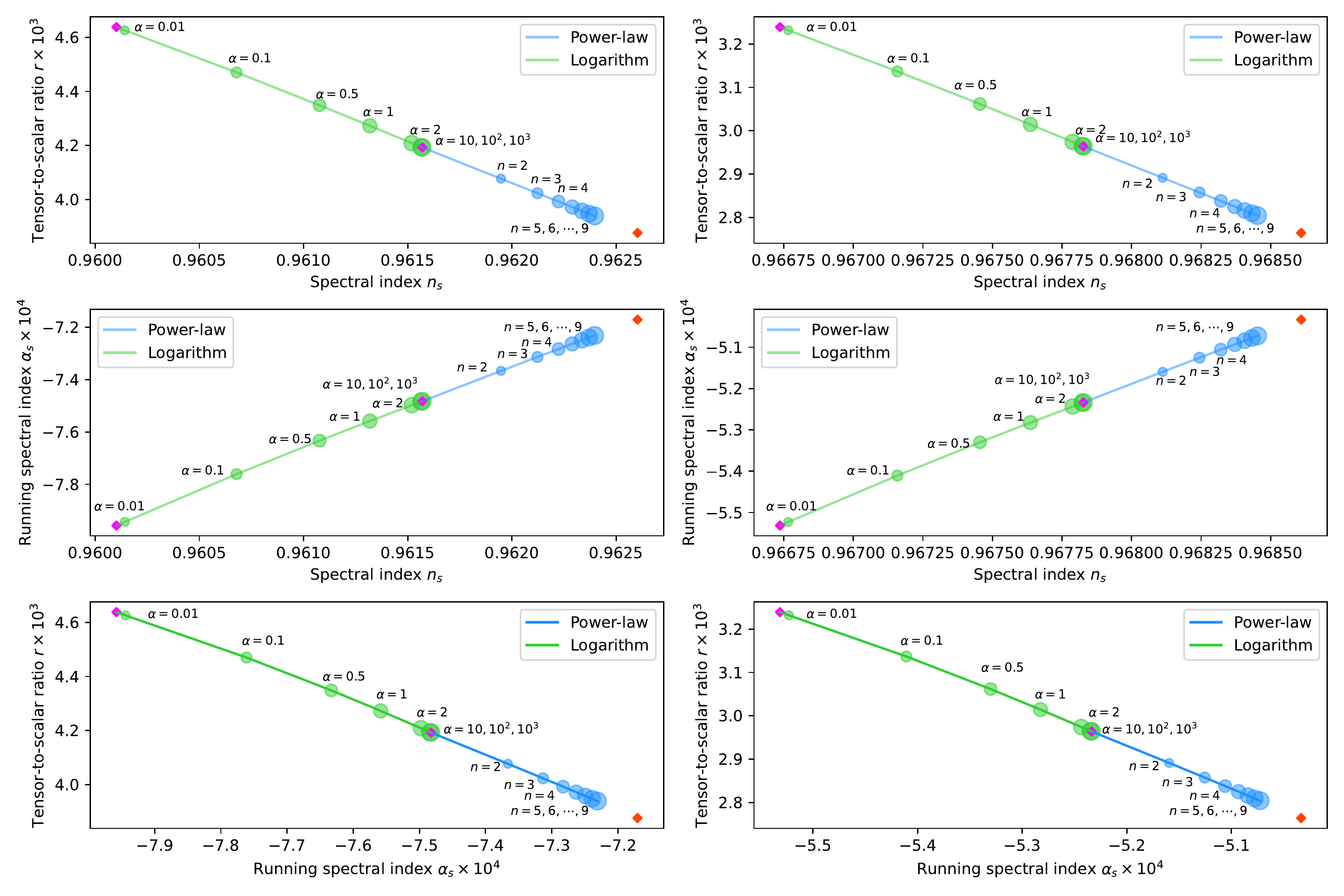}
       \caption{Numerical results of CMB fluctuations, $n_s-r$ (Top), $\alpha_s-r$ (Middle), and $n_s-\alpha_s$ (Bottom) at $N=50$ (Left) and $N=60$ (Right). The Blue lines represent the power-law model with $n=2-9$ and the Green lines are the logarithmic model with $\alpha=0.01,0.1,0.5,1,2,10,100,1000$. The red and magenta diamonds show the attractor points at $\alpha=1/9999$, $\alpha\to\infty$, and $n\to\infty$. }
       \label{Fig::CMBflu}
\end{figure}

First, we consider the power-law model of Eq.~\eqref{Model::PowerLaw}.
The potential of the power-law model can be expressed as
\begin{align}
V(\phi)\propto\left(1-e^{-\sqrt{\frac{2}{3}}\frac{\phi}{{M_ {\rm Pl}}}}\right)^{1+\frac{1}{n-1}}.
\label{Potential::Powerlaw}
\end{align}
At the limit $n\to\infty$, the potential becomes
\begin{align}
\left. V(\phi)\right|_{n\to\infty}\propto 1-e^{-\sqrt{\frac{2}{3}}\frac{\phi}{{M_ {\rm Pl}}}}.
\end{align}
Starting from this potential, we can calculate the CMB fluctuations. 
Figure~\ref{Fig::CMBflu} shows the numerical results for the power-law model with $n=2-9$ and $\infty$.
In these figures, the attractor points at $n\to\infty$ are shown by the red diamonds.
It is remarkable that the results are consistent with the observation even at the limit, $n \to \infty$.
The predicted spectrum indices have a narrow range of variation of about $10^{-3}$ for a change of $n=2$ to $\infty$ in the parameters of the power-law model.
\begin{figure}
\centering
      \includegraphics[width=0.6\linewidth]{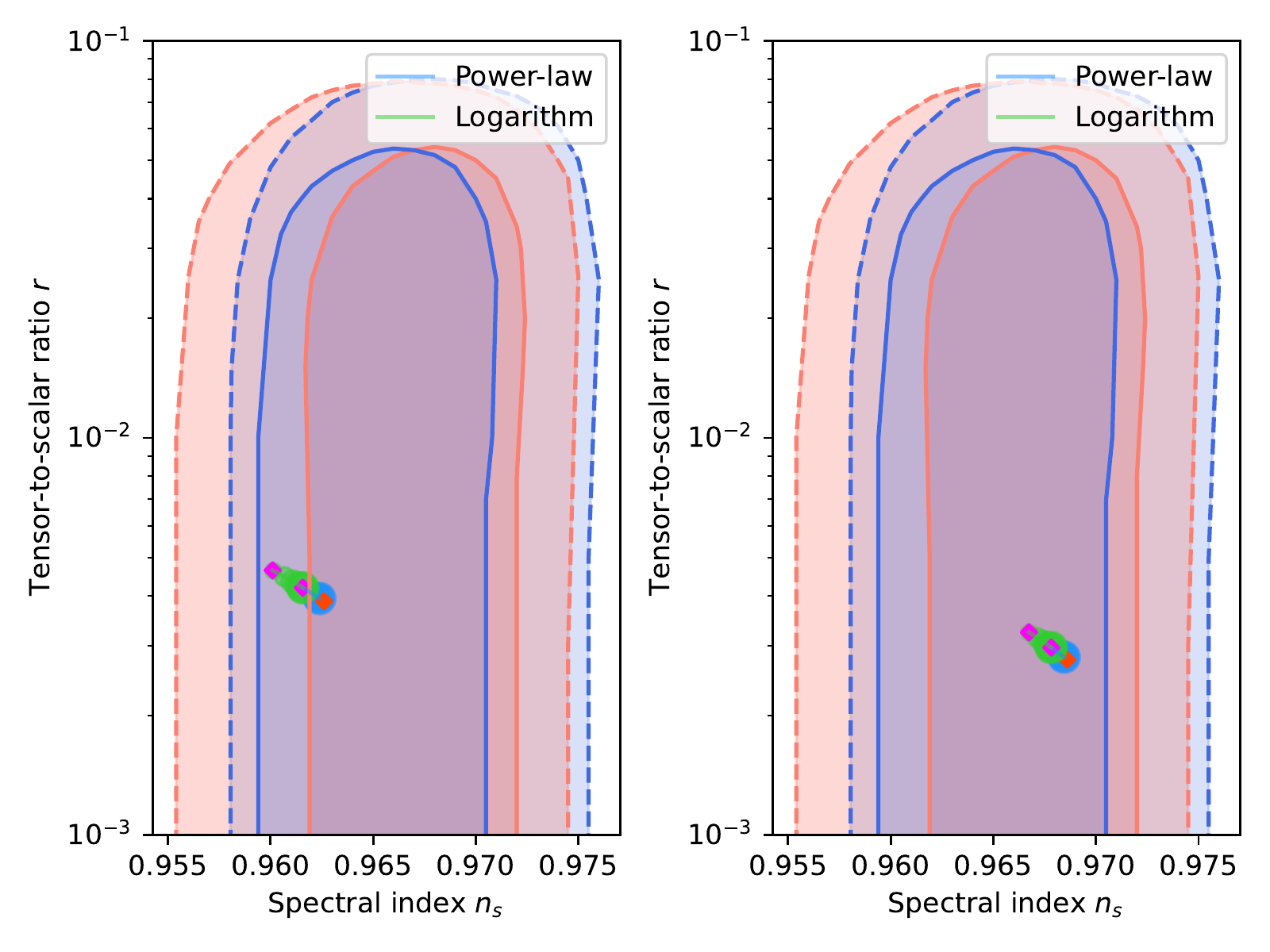}
       \caption{Numerical result of CMB fluctuations, $n_s-r$ for the power-law and logarithmic models at $N=50$ (Left) and $N=60$ (Right) with Planck constraints. The blue and red area shows the Planck 2018 constraints in  \cite{Planck:2018jri}.}
       \label{Fig::CMBflu::Plancl}
\end{figure}

Next, we analyze the logarithmic model~\eqref{Model::Logarithmic} and show the results in Fig.~\ref{Fig::CMBflu}.
We observe that the CMB fluctuations in the logarithmic model approach to those in the $R^2$ model as $\alpha$ increases.
As explained at the end of the previous section, this model is close to the Starobinsky model at the limit, $\alpha \to \infty$.
Thus, we can understand why the fluctuations approach those of the $R^2$ model as $\alpha$ increases.

For small coupling, this model is approximated as $f(R)\sim R^{1+\alpha}$ from Eq.~\eqref{Log::approximation::small}.
Consequently, the logarithmic model with small coupling can be rewritten by the power-law model with $n=1+\alpha$.
From Eq.~\eqref{Potential::Powerlaw} the potential is given by
\begin{align}
V(\phi)\propto\left(1-e^{-\sqrt{\frac{2}{3}}\frac{\phi}{{M_ {\rm Pl}}}}\right)^{1+\frac{1}{\alpha}}.
\end{align}
It should be noted that the slow-roll scenario can not be adapted because of the vanishing potential energy at the limit, $\alpha\to0$.
However, as can be seen in Fig.~\ref{Fig::CMBflu}, even at extreme values such as $\alpha=1/9999$, the CMB fluctuations do not vary significantly and show attractor-like behavior.

Next we summarize the numerical results of the power-law and logarithmic models in Figure~\ref{Fig::CMBflu::Plancl} and Table~\ref{Observables::numerical results}.
From Fig.~\ref{Fig::CMBflu::Plancl}, the numerical results of the entire parameter region for Cartan $F(R)$ gravity satisfies the constraints of Planck 2018~\cite{Planck:2018jri}.
In other words, all the results are consistent with the current observations.
In these models the variation in CMB fluctuations is within a narrow range.
These results demonstrate the robustness of certain Cartan $F(R)$ gravity models.

\begin{table}
    \centering
    \begin{tabular}{cccccc}
    \hline
        Model & Parameter & $N$ & $n_s$  & $\alpha_s$ & $r$ \\ \hline
       \multirow{6}{*}{Power-law\eqref{Model::PowerLaw}} & \multirow{2}{*}{$n=2$} & 50 & 0.9616 & -0.000748 & 0.00419 \\
         & ~ & 60 & 0.9678 & -0.000523 & 0.00296 \\ 
         & \multirow{2}{*}{$n=9$} & 50 & 0.9624 & -0.000723 & 0.00394 \\ 
         & ~ & 60 & 0.9685 & -0.000507 & 0.00280 \\ 
          & \multirow{2}{*}{$n\to\infty$} & 50 & 0.9626 & -0.000717 & 0.00388 \\ 
         & ~ & 60 & 0.9686 & -0.000503 & 0.00276 \\ 
       \multirow{8}{*}{Logarithmic\eqref{Model::Logarithmic}} 
       & \multirow{2}{*}{$\alpha=1/9999$} & 50 & 0.9601 & -0.000796 & 0.00464 \\
        ~ & ~ & 60 & 0.9667 & -0.000553 & 0.00324 \\ 
       & \multirow{2}{*}{$\alpha=0.01$} & 50 & 0.9601 & -0.000796 & 0.00464 \\
        ~ & ~ & 60 & 0.9667 & -0.000553 & 0.00324 \\ 
        ~ & \multirow{2}{*}{$\alpha=1000$} & 50 & 0.9616 & -0.000748 & 0.00419 \\
        ~ & ~ & 60 & 0.9678 & -0.000523 & 0.00296 \\
        ~ & \multirow{2}{*}{$\alpha\to\infty$} & 50 & 0.9616 & -0.000748 & 0.00419 \\
        ~ & ~ & 60 & 0.9678 & -0.000523 & 0.00296 \\
        Constraints~\cite{Planck:2018jri} & - & - & $0.967 \pm 0.004$ & $-0.0042 \pm 0.0067 $ & $< 0.065$ \\ \hline
    \end{tabular}
    \caption{Numerical results of spectral index ($n_s$), running spectral index ($\alpha_s$), tensor-to-scalar ratio ($r$) in power-law and Logarithmic models at e-folding number $N=50, 60$.}
    \label{Observables::numerical results}
\end{table}

\section{Reheating}
\label{section::reheating}
\begin{figure}
    \begin{minipage}{0.45\linewidth}
    \centering 
    \includegraphics[,scale=0.5]{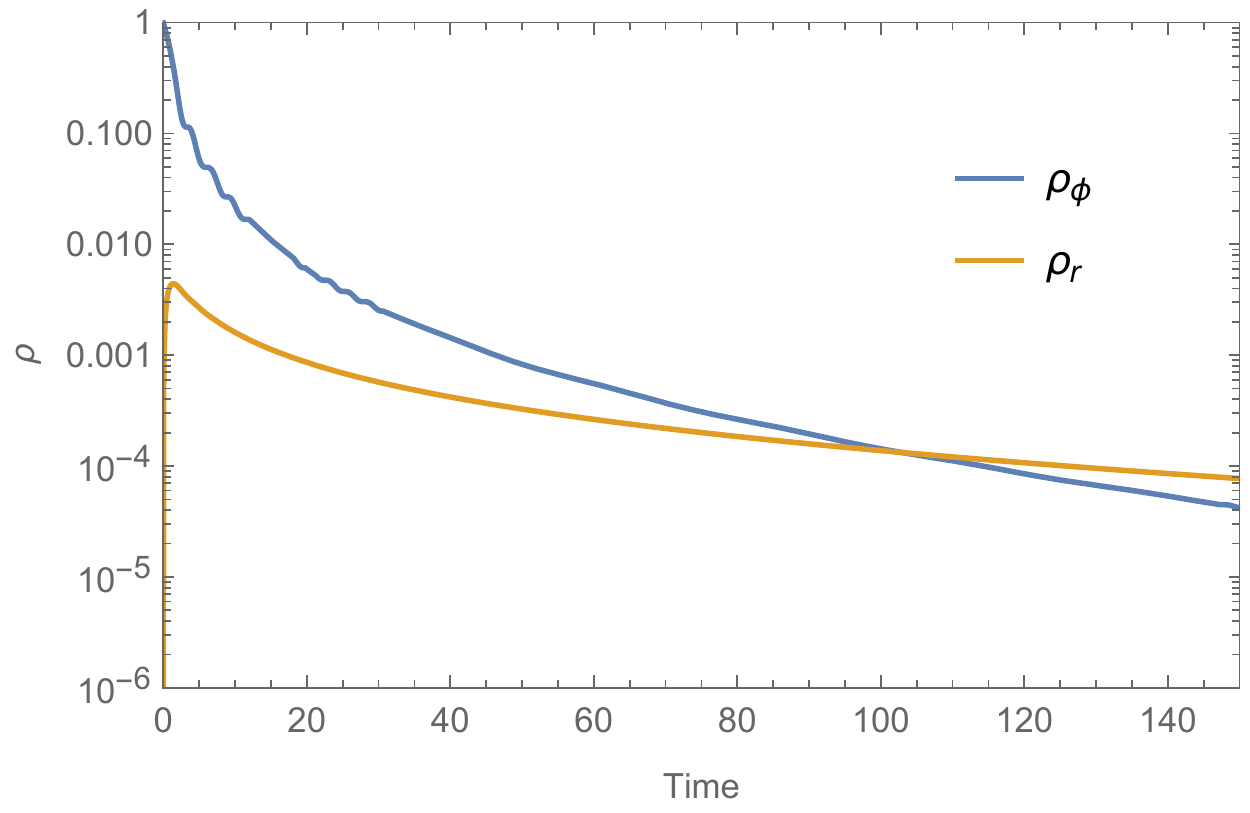}
    \subcaption{Power-law model($n=2$)}
    \end{minipage}
    \begin{minipage}{0.45\linewidth}
    \centering
    \includegraphics[scale=0.5]{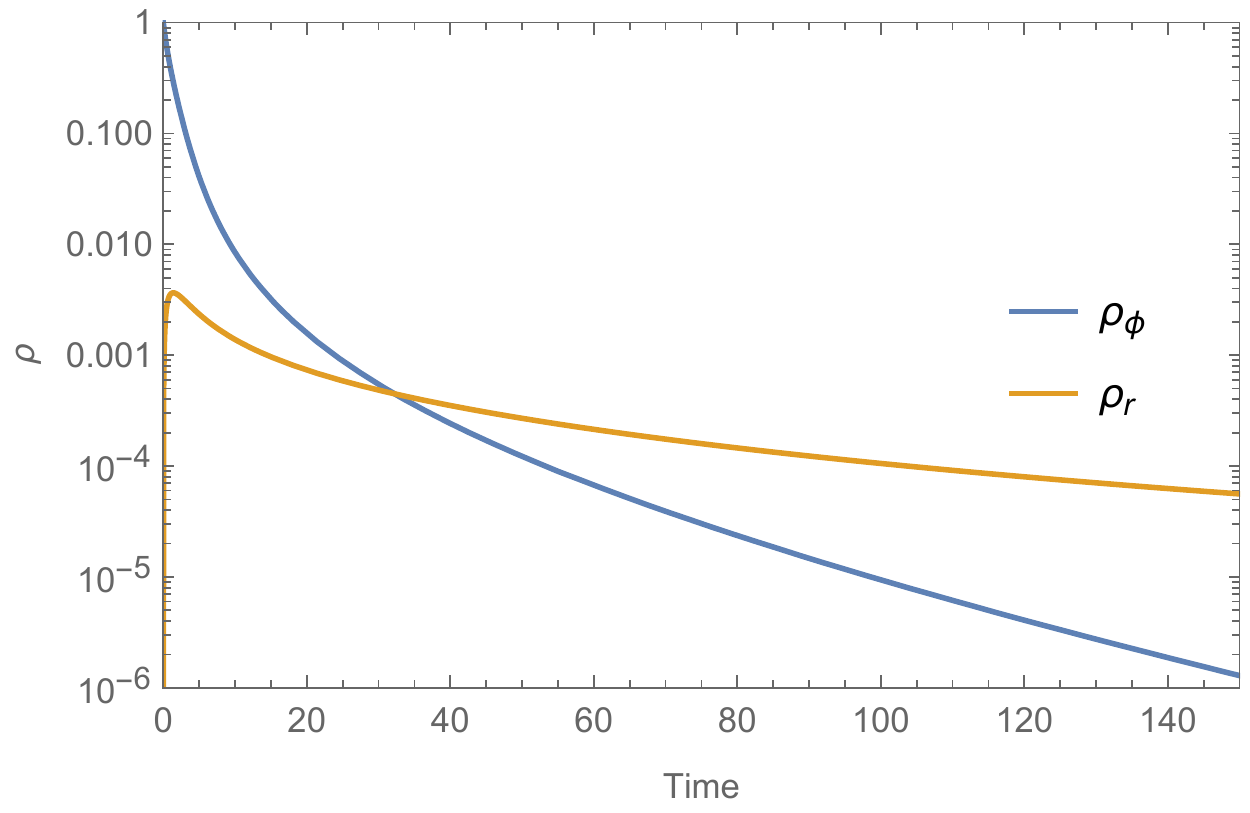}
    \subcaption{Logarithmic model($\alpha=0.01$)}
    \end{minipage}
    \caption{Numerical results of the time evolution of energy densities. Time is scaled by $\tilde{t}=Mt$ and $\Gamma/M=10^{-2}$ where $M=10^{15}$GeV. 
    The initial time is specified at the end of inflation, $\epsilon_V=1$.}\label{Reheating::numerical results::density}
\end{figure}
When considering a realistic cosmological scenario, it is essential to incorporate reheating processes after inflation.
During the reheating process, the energy of the inflaton transitions to radiation and the universe enters a radiation-dominated era.
In this section, we consider the reheating process in Cartan $F(R)$ gravity.

We perform numerical calculations of the reheating process for $n=2$ in the power-law model and $\alpha=0.01$ in the logarithmic model.
Let us assume that the decay rate from inflaton to radiation is $\Gamma$.
The equation of motion for the inflaton becomes
\begin{align}\label{Reheating::phi}
  \ddot{\phi}+3H\dot{\phi}+V'(\phi)=-\Gamma\dot{\phi}. 
\end{align}
Next, the equation for the energy density of radiation is given as
\begin{align}\label{Reheating::radiation}
    \dot{\rho}_r+4H\rho_r=\Gamma\rho_r.
\end{align}
Also, Friedmann equation is 
\begin{align}\label{Reheating::Friedmann}
    3H^2{M_{\rm Pl}}^2 =\rho_\phi +\rho_r,
\end{align}
where $\rho_\phi$ is the energy density of inflaton $\phi$;
\begin{align*}
 \rho_\phi=\frac{1}{2}\dot{\phi}^2+V(\phi).
\end{align*}
From Eqs.~\eqref{Reheating::phi}, \eqref{Reheating::radiation}, and \eqref{Reheating::Friedmann}, the time evolution of the energy densities is calculated numerically.
These results are in Fig.~\ref{Reheating::numerical results::density}.
In addition, Table.~\ref{Reheating::numerical results} shows the starting time of the radiation-dominated era $\tilde{t}_R$ and the reheating temperature $T_R$.
The numerical reheating temperatures are in order agreement with analytical results~\cite{Lozanov:2019jxc}.
We have demonstrated that the reheating process occurs in each model of Cartan $F(R)$ gravity.

For other model parameters, the logarithmic model with a large coupling approximates the $R^2$ model with the Starobinsky potential.
In the power-law model with $n>2$, the potential can be approximated as $|\phi|^{1+\frac{1}{n-1}}$ around $\phi=0$ when $n$ is even.
Although there may be a slight difference in the oscillation at the bottom of the potential, a similar reheating process can be considered.
On the other hand, when $n$ is odd, a region where $\phi<0$ cannot be defined, and instant reheating or preheating is required~\cite{Felder:1998vq,deHaro:2023xcc}.

In this study, we assume a constant friction term, $\Gamma$.
As a future development, we aim to consider the friction term through interactions obtained from Cartan formalism.
\renewcommand{\arraystretch}{1.2}
\begin{table}
    \centering
    \begin{tabular}{ccc}
    \hline
    Model                 & $\tilde{t}_R$  & $T_R$[GeV]  \\ 
        \hline
    Power-law($n=2$)     & 103 &$5.4\times10^{10}$\\
    Logarithmic($\alpha=0.01$)& 32.2&$7.4\times10^{10}$
    \\\hline
    \end{tabular}
    \caption{Numerical results of reheating in power-law and Logarithmic models. \label{Reheating::numerical results}}   
\end{table}

\section{Conclusion}
\label{section::conclusion}
We have studied Cartan $F(R)$ gravity, an extension of $F(R)$ gravity on Riemann-Cartan geometry.
We constructed a derivation of the scalar-tensor theory from Cartan $F(R)$ gravity.
A scalar field with a canonical kinetic term is introduced by extracting the torsion from the curvature scalar.
The potential term is derived from the modified gravity action, $f(R)$.
Since the derivation does not require a conformal transformation, it is free from the equivalence problem between Jordan and Einstein frames in conventional $F(R)$ gravity.

The derived scalar-tensor theory has been applied to the slow-roll scenario of inflation.
We have developed the formulations in Eqs.~\eqref{epsilonv::slow-roll}-\eqref{xiv::slow-roll} and \eqref{eFoldingNumber::generalForm} for the CMB fluctuations in Cartan $F(R)$ gravity. 
In these formulations, it is possible to compute the results directly from the $F(R)$ form without expressing the potential in terms of a scalar field.
The CMB fluctuations have been calculated for the power-law and logarithmic models.
We have found that the obtained results are consistent with observations and concluded that Cartan $F(R)$ gravity gives a realistic inflation model.
We have also shown that the CMB fluctuations are robust to variations in the model parameters.
Additionally, this means that Cartan $F(R)$ gravity is also valid when a low-energy effective theory of quantum gravity is assumed to present a polynomial form.
These results in the power-law model differ from the conventional $F(R)$ gravity where the potential has a local maximum if the exponent is larger than two ($n>2$) and fine-tuning is unavoidable to have a realistic e-folding number~\cite{Inagaki:2019hmm}.  

It is interesting to investigate whether the robustness of Cartan $F(R)$ gravity is a generic feature.
We would like to apply the model for the reheating process after inflation~\cite{Nishizawa:2014zra,Oikonomou:2017bjx,Mathew:2020nqo,Rajabi:2022qrs}. 
The reheating process may also reveal different features from the conventional $F(R)$ gravity.
The original ECKS theory has a four-fermion interaction called spin-spin interaction or Dirac-Heisenberg-Ivanenko-Hehl-Datta four-body fermi interaction~\cite{Hehl:1974cn,Kerlick:1975tr,Gasperini:1986mv,Hehl:1971qi,Boos:2016cey}.
Cartan $F(R)$ gravity has been associated with matter fields such as spin-spin interaction through torsion.
The interaction between the inflaton and matter fields produces a reheating process.
It then reveals the growth of the universe leading to standard cosmology.

\section*{Acknowledgements}
For valuable discussions, the authors would like to thank N.~Yoshioka.
This work was supported by JST, the establishment of university fellowships towards the creation of science technology innovation, Grant Number JPMJFS2129.

\appendix

\section{Lambert's function}
\begin{figure}
    \centering
    \includegraphics[width=0.7\linewidth]{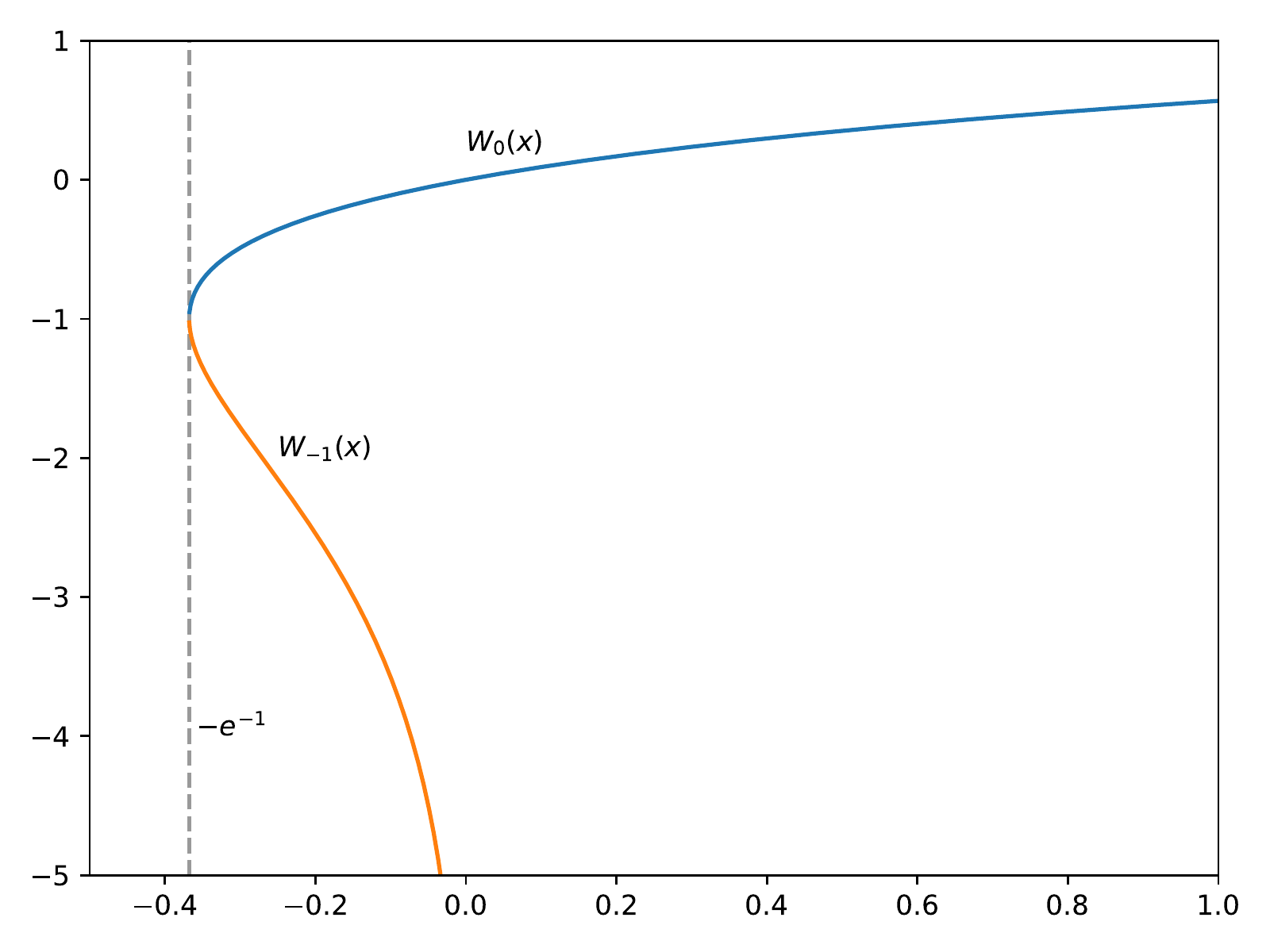}
    \caption{
        Two branches of the Lambert's $W$ function.
        The blue line, $W_0$, is the principal branch which is defined the interval $(-e^{^1}, \infty)$
        and the red line, $W_{-1}$, is other branch in the interval $(-e^{-1}, 0)$.
        The branching point is at $(-e^{-1}, -1)$.
    }
    \label{LambertFunction::Fig}
\end{figure}
In this section, we briefly introduce Lambert's $W$ function.
For more details see Ref.~\cite{DBLP:journals/corr/abs-1003-1628} as an example.
The function $W(x)$ is defined to satisfy the following equation
\begin{align}
    W(x) e^{W(x)} = x,
    \label{LambertFunction::Definition}
\end{align}
where $W(x)$ is called the Lambert's $W$ function.
The $W$ function has two branches. One is defined in the interval $[-e^{-1}, \infty]$, and other is in the interval $(-e^{-1}, 0)$.
The former is called the principal branch and described as $W_0$,
while the latter is written as $W_{-1}$.
The branching point is at $(-e^{-1}, -1)$.
See, fig.~\ref{LambertFunction::Fig}.

For $x > 0$, the principal branch can rearrange eq.~\eqref{LambertFunction::Definition} to take the natural log,
\begin{align}
    W(x) = \ln x - \ln W(x),
\end{align}
then we can obtain the recursive relation,
\begin{align}
    W_0(x) = \ln x - \ln (\ln x - \ln (\ln x - \cdots)),
    \label{LambertFunction::Principal::Recursive}
\end{align}
and, for $x < 0$, we can obtain the similar relation for $W_{-1}$ as
\begin{align}
    W_{-1}(x) = \ln(-x) - \ln (-(\ln(-x) - \ln (-\ln (-x + \cdots)))).
    \label{LambertFunction::Negative::Recursive}
\end{align}

The asymptotic expansion of Lambert function is estimated by
using Lagrange inverse theorem as
\begin{align} \nonumber
  W(x) =& L_1 - L_2 + \frac{L_2}{L_1} + \frac{L_2 (-2 + L_2)}{2L_1^2}
  + \frac{L_2 (6 - 9L_2 + 2L_2^2)}{6L_1^3}
  \\ &+ \frac{L_2 (-12 + 36L_2 - 22L_2^2 + 3L_2^3)}{12L_1^4}
  + \mathcal{O}\Big(\big\{\frac{L_2}{L_1}\big\}^5\Big),
  \label{LambertFunction::AsymptoticExpansion}
\end{align}
where $L_1 = \ln x$ and $L_2 = \ln\ln x$ for the principle branch,
and $L_1 = \ln(-x)$ and $L_2 = \ln(-\ln(-x))$ for the negative branch.
Note that the expansion~\eqref{LambertFunction::AsymptoticExpansion} is convergent
at both 0 and infinity. See~\cite{Corless1996} for details.

\section{Exact solution of Starobinsky model in Riemann geometry}

As the power-law model of Cartan $F(R)$ gravity,
the inflationary observables of the
Starobinsky model in Riemann geometry can be
represented analytically.

The e-folding number of Starobinsky model is expressed by
\begin{align}
    N = \frac{3}{4} \Big(
        2 \gamma R - \ln (1 + 2 \gamma R)
    \Big) \Big|_{R_\text{end}}^{R_*},
\end{align}
and the curvature $R_*$ is solved as
\begin{align}
    R_* = -1 - W_{-1}\Big(-e^{-\big(1 + 4N/3\big)}\Big).
\end{align}

\begin{align}
    n_s = 1 + \frac{8}{3} \frac{1}{1 + W_{-1}(-e^{-(1+4N/3)})}
    - \frac{16}{3} \Big(\frac{1}{1 + W_{-1}(-e^{-(1+4N/3)})}\Big)^2
    \sim 1 - \frac{2}{N} - \frac{3}{N^2},
    \label{Starobinsky::spectrumIndex}
\end{align}

\begin{align}
    r = \frac{64}{3} \Big(\frac{1}{1 + W_{-1}(-e^{-(1+4N/3)})}\Big)^2
    \sim \frac{12}{N^2},
    \label{Starobinsky::tensorScalarRatio}
\end{align}

\begin{align}
  \begin{aligned}
    \alpha_s =& -\frac{32}{9} \Big(\frac{1}{1 + W_{-1}(-e^{-(1+4N/3)})}\Big)^2
    + \frac{160}{9} \Big(\frac{1}{1 + W_{-1}(-e^{-(1+4N/3)})}\Big)^3
   \\
   &- \frac{128}{3} \Big(\frac{1}{1 + W_{-1}(-e^{-(1+4N/3)})}\Big)^4
    \\ \sim& -\frac{2}{N^2} - \frac{15}{2N^3} - \frac{27}{2N^4}.
  \end{aligned}
  \label{Starobinsky::runningSpectrumIndex}
\end{align}
The approximation of r.h.s of eqs.~\eqref{Starobinsky::spectrumIndex}, \eqref{Starobinsky::tensorScalarRatio}, \eqref{Starobinsky::runningSpectrumIndex}
are evaluated to apply the recursive formula~\eqref{LambertFunction::Negative::Recursive}.

\bibliography{ref}
\end{document}